\documentstyle[aps,epsfig]{revtex}

\newcommand{\Ref}[1]{(\ref{#1})}
\newcommand{\p}{\partial}
\begin{document}

\twocolumn[\hsize\textwidth\columnwidth\hsize\csname
@twocolumnfalse\endcsname

\title{\Large\bf Reconcile Planck-scale
discreteness \\ and the Lorentz-Fitzgerald contraction}

\author{Carlo Rovelli {\it ${}^{ab}$}, Simone Speziale
{\it ${}^{b}$} \\[1mm]
\it ${}^a$  Centre de Physique Th\'eorique de Luminy,
Case 907, F-13288 Marseille, EU\\
\it ${}^b$  Dipartimento di Fisica, Universit\`a di Roma ``La
Sapienza",
I-00185 Roma, EU}

\date{\small\today} \maketitle

\begin{abstract}

    A Planck-scale minimal observable length appears in many
    approaches to quantum gravity.  It is sometimes argued that this
    minimal length might conflict with Lorentz invariance, because a
    boosted observer could see the minimal length further Lorentz
    contracted.  We show that this is not the case within loop quantum
    gravity.  In loop quantum gravity the minimal length (more
    precisely, minimal area) does not appear as a fixed property of
    geometry, but rather as the minimal (nonzero) eigenvalue of a
    quantum observable.  The boosted observer can see the same
    observable spectrum, with the same minimal area.  What changes
    continuously in the boost transformation is not the value of the
    minimal length: it is the probability distribution of seeing one
    or the other of the discrete eigenvalues of the area.  We discuss
    several difficulties associated with boosts and area measurement
    in quantum gravity.  We compute the transformation of the area
    operator under a local boost, propose an explicit expression for
    the generator of local boosts and give the conditions under which
    its action is unitary.
\end{abstract}
\vskip1cm]

\section{Introduction}

A large number of convincing semiclassical considerations indicate
that in a quantum theory of gravity the Planck length $L_{P}$ should
play the role of minimal observable length \cite{minimal}.  Indeed,
this happens, in different manners, in most, if not all, current
tentative quantum gravity theories.  It is often argued that the
existence of this minimal length might signal a problem with Lorentz
invariance (for instance, see \cite{lee}).  A Lorentz invariant
quantum theory can easily accommodate a basic observable length (in a
free quantum field theory of a massive scalar field, for instance,
there is the Compton wavelength of the particle), but is a
\emph{minimal} observable length compatible with some form of Lorentz
invariance?  One might argue that length transforms continuously under
a Lorentz transformation, and a minimal length $L_{P}$ is going to get
Lorentz contracted in a boost.  Thus, a boosted observer should see a
Lorentz contracted $L_{P}$, namely a length shorter than the length
claimed to be minimal, leading to a contradiction.

This arguments is certainly simple minded, but it has had large
resonance on quantum gravity research.  The apparent conflict between
Lorentz transformations and Planck scale discreteness, for instance,
is often quoted as one of the motivations for quantum deformations of
the Lorentz symmetry, and the use of quantum groups, or q-deformed
Lorentz algebras, in this context.  Within canonical quantum gravity,
similar arguments have been used to suggest that no state of the
theory can be locally Lorentz invariant, and so on.

In any case, it is clear that an approach to quantum gravity
predicting that an observer ${\cal O}$ observes a minimal length
$L_{P}$ must answer the question whether or not a boosted observer
${\cal O}'$ can observe this length Lorentz contracted.  And whether
or not, in this sense, Planck scale discreteness can be compatible
with some form of local Lorentz invariance.

Here, we show how the apparent conflict between Lorentz
contraction and Planck scale discreteness is resolved in loop
quantum gravity \cite{loop} (for a review and extended
references, see \cite{loop2} and \cite{tom}.)  Within loop
quantum gravity, a minimal length appears characteristically in
the form of a minimal (nonzero) value $A_{0}$ of the area of a
surface \cite{discreteness,discreteness2}. Here we show that in
loop quantum gravity a boosted observer ${\cal O}'$ does not
observe a Lorentz contracted $A_{0}$.  The minimal (nonzero) area
that the boosted observer ${\cal O}'$ can observe is still
$A_{0}$.  We show that Planck scale discreteness is compatible
with a certain implementation of local Lorentz invariance, and we
study the transformation properties of the area operator under an
infinitesimal local boost.

\subsection{The basic idea}

\begin{figure}[t]
  \centering
  \includegraphics[width=2cm]{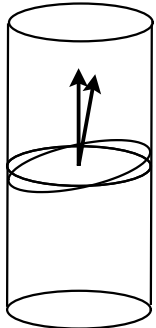}
  \caption{Two observers in relative motion (arrows) see two different
  table's 2d surfaces (ovals) in spacetime, because their
  simultaneity surfaces are different and have thus a different
  intersection with the table worldsheet (cylinder).}\label{1}
\end{figure}

The key to understand how this may happen is the fact that in loop
quantum gravity a minimal length does not appear as a fixed structural
property of space geometry.  Space geometry, indeed, has no fixed
structural property at all in this approach.  The geometry of space
comes from a quantum field, the quantum gravitational field. 
Therefore the observable properties of the geometry, such as, in
particular, a length, or an area, are observable properties of a
quantum physical system.  A measurement of a length is therefore a
measurement in the quantum mechanical sense.  Generically, quantum
theory does not predict an observable value: it predicts a probability
distributions of possible values.  Given a surface moving in
spacetime, the two measurements of its area performed by two observers
${\cal O}$ and ${\cal O}'$ boosted with respect to each other are two
distinct quantum measurements.  Correspondingly, in the theory there
are two distinct operators $A$ and $A'$, associated to these two
measurements.  Now, our main point is the technical observation that
$A$ and $A'$ do not commute:
\begin{equation}
    [A,A'\,]\ne 0.
    \label{eq:nc}
\end{equation}
This is because $A$ and $A'$ depend on the gravitational field on two
distinct 2d surfaces in spacetime (see Figure \ref{1}) and a field
operator does not commute with itself at different times.  In this
paper, we prove equation \Ref{eq:nc}.
It follows that a generic eigenstate of $A$ is not an eigenstate of
$A'$.  If the observer ${\cal O}$ measures the area and obtains the
minimal value $A_{0}$, the state of the gravitational field will be
projected on an eigenstate of $A$.  This, in turn, is not going to be
an eigenstate of $A'$.  If then the observer ${\cal O}'$ measures the
area, he will therefore find the state in a superposition of
eigenstates of $A'$.  That is to say, the theory predicts that, for
him, the surface does not have a sharp area.  If the experiment is
repeated several times, ${\cal O}'$ will observe a probability
distribution of area values.  The mean value of the area can be
Lorentz contracted, while the minimal nonzero value of the area can
remain $A_{0}$.

The situation is analogous to what happens with angular momentum in
the ordinary quantum mechanics of a rotationally invariant system with
given (say half-integer) spin.  Consider a certain direction, say the
$z$ direction.  If we measure the component $L_{z}$ of the angular
momentum, we have a discrete spectrum with a minimal nonzero value
$L_{0}$.  One might argue that this prediction conflicts with rotation
invariance: if classical angular momentum components change
continuously under a rotation -- how can then an angular momentum
component have a minimal value?  But of course this concern is ill
founded.  If an observer ${\cal O}'$ rotated with respect to $\cal O$
observes his own angular momentum component $L_{z}'$, he will still
observe the same minimal (nonzero) value $L_{0}$.  In particular, if
the observation follows the observation of the value $L_{0}$ by ${\cal
O}$, and if the experiment is repeated, ${\cal O}$ will observe a
distribution of eigenvalues which is uniquely determined by the well
known representation theory of the rotation group in the Hilbert space
of the theory.  The same, we argue here, happens with the area in loop
quantum gravity.

Although this analogy is very illuminating, however, the quantum
gravity situation is far more complicated, for a number of reasons:

\begin{description}
    \item[i.]  The theory as a whole is not Lorentz invariant, and a
    form of Lorentz invariance can only be recovered locally and/or in
    certain (``sufficiently flat") regimes.  \item[ii.]  The area $A$
    is a far more complicated function of the basic variables of the
    theory than $L_{z}$.  \item[iii.]  Lorentz transformations, unlike
    rotational symmetry, do not happen at fixed time.  Therefore the
    generators of the (local) Lorentz transformations have to know
    about the dynamics of the theory, which is highly nontrivial in
    quantum gravity.  \item[iv.]  The very construction of the
    ``Lorentz rotated" quantity $A'$ is delicate, since it involves a
    careful analysis in a \emph{general} relativistic context of what
    it means to measure the area of a surface for a boosted observer.
    \item[v.]  The theory is invariant under diffeomorphisms; the area
    of a surface defined by coordinate values is not gauge invariant
    and we need a physical dynamical quantity to fix the surface whose
    area we want to consider \cite{Rovelli:1990ph}.
\end{description}
For all these reasons, it is not obvious that the quantum area can
behave ``as the $L_{z}$ component of the angular momentum".  In this
paper, we analyze all these problems with care, and we show that in
spite of all these complications, and under certain reasonable
assumptions, what happens to the area under a Lorentz boost in loop
quantum gravity is indeed precisely what is described above and is
illustrated by the analogy with the angular momentum.

Our strategy is the following.  First, we address ({\bf v.}) by
considering a physical system formed by general relativity coupled to
a minimal and realistic amount of matter, sufficient to have a well
defined and diffeomorphism invariant notion of area.  Notice that this
is precisely the context in which the claim that the discretization of
the area is a physically observable prediction of the theory was put
forward \cite{Rovelli:1992vv}.  Second, we address ({\bf iv.}) by
carefully discussing the meaning of the measurement of the area $A'$
``seen" by a boosted observer in classical general relativity (Section
\ref{classical}.)  Then, we solve ({\bf ii.}) by explicitly computing
$A$ and $A'$ as functions of the canonical variables of the theory
(Section \ref{geometry}.)  This is done in a power expansion in the
boost parameter, which allows us to address ({\bf iii.}) by expressing
quantities at $t>0$ in terms of quantities at $t=0$, using the
equations of motion.  In turn, this result allows us to derive
(\ref{eq:nc}) and compute explicitly the first terms of this
commutator in an expansion in the boost parameter (Section
\ref{incomp}.)  Then (Section \ref{quantum}), we construct a quantity
that we suggest could generate the boost.  This generator depends on
the hamiltonian constraints, thus addressing ({\bf iii.}).  Finally in
Section \ref{unitarity} we derive the conditions under which this
transformation is unitary, and thus the spectrum preserved.

Finally, ({\bf i.}) is addressed by means of a delicate interplay
between the full dynamical structure of the theory and the request of
local flatness needed to have Lorentz invariance over a small
spacetime region.  We are interested in small scale quantum
discreteness and small scale quantum fluctuations of the gravitational
field, in quantum states in which the metric is macroscopically flat. 
That is, in which the macroscopic expectation value of the metric
operator, is flat.  To describe this regime, we first analyze the
problem in the classical theory: we expand for small boost parameter
and small surface, and keep only the lowest order relevant terms.  We
then assume that in the quantum theory the expansion remains valid in
the regimes where the expectation value of the macroscopic curvature
is small.  This is not different from what we usually do in
conventional quantum field theory: we take the field to be zero in the
vacuum and expand around this value -- even if the field fluctuates
widely on small scale, and its value is moved far away from zero by a
field measurement at small scale.  Of course in nonperturbative
quantum gravity we have far less control on the quantum state of the
gravitational field that corresponds to macroscopical flat space, and
therefore the viability of this approach we should cautionally be
regarded as an hypothesis.

In addition, in Section \ref{ac} we briefly discuss an alternative
point of view, which we have learned in conversations with
Amelino-Camelia, on the non commutativity between $A$ and $A'$.  The
idea is to view the non commutativity of $A$ and $A'$ as a consequence
of the noncommutativity between the area of the surface and the
relative velocity of the observer and the surface.  We refer to
\cite{giovanni} for a more extensive discussion.

\section{Geometry}
\label{classical}

\subsection{The system}

We consider the physical system formed by four physical elements:
\begin{description}
    \item[i.] the gravitational field,
    \item[ii.] two particles,
    \item[iii.] a two-dimensional surface (the ``table").
\end{description}
These are the dynamical quantities of the system we consider.  They
provide a minimal setting in which we can compare the area observed by
two observers boosted with respect each other.  We are interested in
the area of the table, as seen by two observers (${\cal O}$ and ${\cal
O}'$), moving with the two particles.

Besides these dynamical quantities, we assume that all sort of other
physical objects exist in the universe. These  can be used to perform
measurements (for instance, light pulses traveling along geodesics,
apparatus that detect the arrival of these light pulses, clocks that
measure proper time along world lines, recording devices and so on).
We do not consider these other physical objects as part of the
dynamical system observed: we consider them as part of the measuring
apparatus.  To be precise, we assume that the well known freedom
of choosing the boundary between the observed quantum system and the
classical apparatus --emphasized by Von Neumann-- allows us to do so 
in
this context.

We describe the system in a general relativistic setting as
follows.  We consider a 4d manifold $\cal M$, with coordinates
$x^\mu$, on which the following quantities are defined:
\begin{description}
\item[i.]  The gravitational field $g$ is described by the metric
tensor $g_{\mu\nu}(x)$.  \item[ii.]  The world lines $X$ and $X'$
of the two observers are given by the functions
\begin{eqnarray}
X&:&R\to{\cal M} \nonumber \\
&:& \tau\mapsto x^\mu(\tau)
\end{eqnarray}
and
\begin{eqnarray}
X'&:&R\to{\cal M} \nonumber \\
&:& \tau'\mapsto x^\mu(\tau')
\end{eqnarray}
(we follow here the bad physicists' habit of indicating functions with
the name of the independent and dependent variable: $x^\mu(\tau')$ is
given by a \emph{different function} that $x^\mu(\tau)$, of course.)
\item[iii.]  The world sheet $T$ of the table is described by the
three-dimensional hypersurface
\begin{eqnarray}
T&:& [-1,+1]\times[-1,+1]\times R\to{\cal M} \nonumber \\
&:& (\tau^1,\tau^2,\tau^3)\mapsto x^{\mu}(\tau^1,\tau^2,\tau^3)
\end{eqnarray}
\end{description}
The functions $(g_{\mu\nu}(x),\ x^\mu(\tau),\ x^\mu(\tau'), \
x^{\mu}(\tau^1,\tau^2,\tau^3))$ are the lagrangian variables of
the system.  We assume the dynamics of this system to be governed by
the the Einstein equations and the dynamical equations of the table
and the particles.  For simplicity, we assume that the matter
energy-momentum tensor is negligible in the Einstein equations, but
this is not essential in what follows.

We are interested in a specific subset of physical configurations.
First, we want the world lines of the two observers to cross at a 
point
$P$ situated on the table world sheet.  Second, (in the classical
analysis) we assume that the curvature at and around $P$ and the
acceleration of the particles at $P$ are negligible at the scale of
the surface.  That is, we take the surface small enough that we can
expand around $P$ and keep the lowest terms only.

What is the area $A$ of the table seen by $\cal O$ when at $P$?  The
answer is the following.  $A$ is the area of the 2d surface $S$ formed
by the intersection of the 3d table's worldsheet $T$ with the 3d
simultaneity surface $\Sigma$ of $\cal O$ at $P$.

\begin{figure}[t]
  \centering
  \includegraphics[width=5cm]{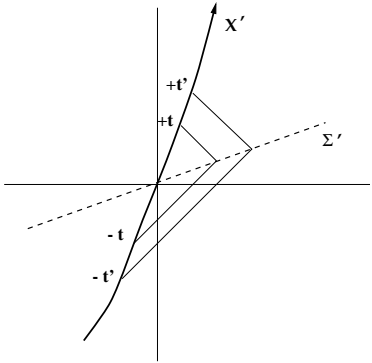}
  \caption{The definition of the simultaneity surface.}\label{2}
\end{figure}

The simultaneity surface $\Sigma$ is the set of points in $\cal M$
whose light cone intersects $X$ in two points at the same proper time
distance (along $X$) from $P$.  Physically, these are the events where
a mirror reflects a light pulse emitted by the observer at proper time
$-t$ such that the reflected pulse gets back to the observer at proper
time $+t$ ($t=0$ being at $P$).  This is Einstein definition of
(relative) simultaneity. (See Figure \ref{2}.)

\begin{figure}[t]
  \centering
  \includegraphics[width=5cm]{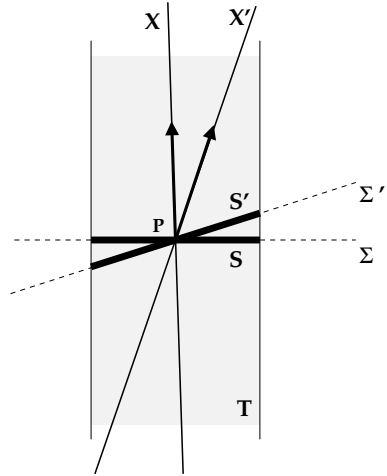}
  \caption{The definition of $S$ and $S'$.}\label{3}
\end{figure}

The intersection between the surface of simultaneity of the observer
$\Sigma$ and the table world history $T$ is a two dimensional surface
$S=\Sigma \cap T$.  It represents the ``table at fixed time" in the
frame of the observer $\cal O$ at $P$.  The area $A$ is the integral
over $S$ of the determinant of the restriction ${}^2g$ of the metric
$g$ to $S$.  The area $A$ is therefore a complicated function
$A[g,X,X',T]$ of ---but it is completely determined by--- the metric 
$g$
the world line $X$, the hypersurface $T$, and the crossing point 
$P$.  We
calculate this function explicitly in Section \ref{agr}.  Similarly,
$A'$ is the area of the intersection between $T$ and the simultaneity
surface of ${\cal O}'$ at $P$.  (See Figure \ref{3}.)

We call $v^\mu$ (and, respectively, $v'{}^\mu$) the unnormalized
four-velocity of $X$ (respectively $X'$) at $P$
\begin{eqnarray}
    v^\mu  = \left.\frac{dx^\mu(\tau)}{d\tau}\right|_{P}.
\end{eqnarray}
($v'{}^\mu$ is defined in the same manner by $x^\mu(\tau')$.)  The
angle between the two tangents gives the relative speed $V$ of the two
observers ${\cal O}$ and ${\cal O}'$
\begin{eqnarray}
\gamma=\frac{1}{\sqrt{1-V^2}}=
\frac{v \cdot v'{}}{|v|\, |v'|}.
\end{eqnarray}
where the scalar product and the norm are taken here with the metric
at $P$.  For simplicity, we also assume that the relative
three-velocity of the two observers is tangent to the table.  (We are
not interested in transversal motion because it does not give rise to
Lorentz contraction.)  We say that the table is at rest with respect
to $X$, if the world sheet of the boundary of the table is normal to
$\Sigma$.  (If the surface is sufficiently small this implies that $A$
maximizes the area with respect to $v$.)

The quantities $A$ and $A'$ are diffeomorphism invariant functions of
$g$, $X$, $X'$ an $T$.  They are invariant under a smooth displacement
of these dynamical quantities on $\cal M$.  They do not depend on the
coordinates chosen on $\cal M$, nor on any structure on $\cal M$
besides the dynamical fields.  They are fully gauge invariant
observables in this dynamical system.  They are physical quantities
that are in principle observable, by using appropriate measuring
devices (formed by light pulses, detectors, clocks, and else).  The
specific technical construction of these devices is not relevant
here.

In this paper we consider the quantum theory corresponding to this
dynamical system.  In particular, we consider the quantum operators
corresponding to the physically observable quantities $A$ and $A'$, we
show that (\ref{eq:nc}) is true, and that the operator $A'$ can be
obtained (under certain assumptions) from a unitary transformation
that implements a local Lorentz transformation in the Hilbert space of
the theory.

\subsection{Area in general relativity}\label{agr}

What is the area $A(S)$ of the (small) 2d surface $S$ given by the
intersection of two 3d hypersurfaces $\Sigma$ and $T$?  Here we show
that $A(S)$ can be written in terms of the one-forms $n_{\mu}^\Sigma$
and $n_{\mu}^T$ normal to the two hypersurfaces.  We shall then use
this fact to directly connect the area to the motion of the
observers.  The table worldsheet $T$ is parametrized by
$x^{\mu}(\tau^1,\tau^2,\tau^3)$.  Its normal one-form is
\begin{eqnarray}
n^T_{\mu}=\epsilon_{\nu\rho\sigma\mu}\ \frac{\p x^{\nu}}{\p \tau^1
}\frac{\p x^{\rho}}{\p \tau^2 }\frac{\p x^{\sigma}}{\p \tau^3 }.
\end{eqnarray}
It does not depend on the metric.
Similarly, the normal of the hypersurface $\Sigma$,
parametrized as $x^{\mu}(\rho^1,\rho^2,\rho^3)$, is
\begin{eqnarray}
\label{normale}
n^{\Sigma}_{\mu}=\epsilon_{\nu\rho\sigma\mu}\frac{\p x^{\nu}}{\p
\rho^1 }\frac{\p x^{\rho}}{\p \rho^2 }\frac{\p x^{\sigma}}{\p
\rho^3 }.
\end{eqnarray}
The normal two-form of the intersection
$S=\Sigma \cap T$ parametrized as $x^\mu(u,v)$, is
\begin{eqnarray}
n_{\mu\nu}=\frac{1}{2}\epsilon_{\mu\nu\rho\sigma}\frac{\p x^{\rho}}{\p
u}\frac{\p x^{\sigma}}{\p v }.
\end{eqnarray}
It is convenient to choose parametrizations such that
\begin{eqnarray}
u= &\tau^1& = \rho^1 \nonumber \\
v= &\tau^2& = \rho^2
\end{eqnarray}
and
\begin{eqnarray}
\epsilon_{\mu\nu\rho\sigma}\
\frac{\p x^{\mu}}{\p u}
\frac{\p x^{\nu}}{\p v}
\frac{\p x^{\rho}}{\p \tau^3 }
\frac{\p x^{\sigma}}{\p \rho^3 }=1.
\end{eqnarray}
Then we have easily
\begin{eqnarray}
n_{\mu\nu} =\ n^\Sigma_{[\mu}\ n^T_{\nu]}.
\end{eqnarray}
The area of a 2d surface is
\begin{eqnarray}
A =
A(S)&=&\int_Sdudv\ \sqrt{\det {}^2g} \nonumber \\    &=&\int_S
dudv\ \sqrt{\det \left(\frac{\p x^\mu }{\p
u^i}\frac{\p  x^\nu}{\p u^j}\ g_{\mu\nu} \right)}.
\end{eqnarray}
where $u^i=(u,v)$ and the determinant is on the $i,j=1,2$ indices.

Consider now the equality
\begin{eqnarray}
2 gn_{\mu\nu}n_{\alpha\beta}g^{\mu\alpha}g^{\nu\beta}&=&
\frac{g}{2}\epsilon_{\mu\nu\rho\sigma}
\epsilon_{\alpha\beta\gamma\delta}\frac{\p x^{\rho}}{\p u
}\frac{\p x^{\sigma}}{\p v  }\frac{\p x^{\gamma}}{\p u
}\frac{\p x^{\delta}}{\p v  }g^{\mu\alpha}g^{\nu\beta}\nonumber \\    
&=&
-\frac{1}{2} \epsilon_{\mu\nu\rho\sigma}\epsilon^{\mu\nu\zeta\theta}
\frac{\p x^{\rho}}{\p u  }\frac{\p x^{\sigma}}{\p v
}\frac{\p x_{\zeta}}{\p u  }\frac{\p x_{\theta}}{\p v
}\nonumber \\    &=& \frac{\p x^{\rho}}{\p u  }\frac{\p x_{\rho}}{\p
u  }\frac{\p x^{\sigma}}{\p v  }\frac{\p x_{\sigma}}{\p
v  }-\left(\frac{\p x^{\rho}}{\p u  }\frac{\p x_{\rho}}{\p
v  }\right)^2 \nonumber \\    &=&\det \left(\frac{\p x^\mu }{\p
u^i}\frac{\p  x^\nu}{\p u^j}\ g_{\mu\nu}\right),
\end{eqnarray}
obtained using
\begin{eqnarray}
\epsilon^{\alpha\beta\gamma\delta}&=&-g\epsilon_{\mu\nu\rho\sigma}
g^{\mu\alpha}g^{\nu\beta}g^{\gamma\rho}g^{\delta\sigma},  \nonumber \\
\epsilon_{\mu\nu\rho\sigma} \epsilon^{\mu\nu\zeta\theta}&=&2\
(\delta^\zeta_\sigma\delta^\theta_\rho-
\delta^\zeta_\rho\delta^\theta_\sigma).
\end{eqnarray}
Using this, the area of $S$ can be written as
($g= \det\ g_{\mu\nu})$
\begin{eqnarray}
\label{areaold}
A &=&\int_Sdu dv \ \sqrt{2\ g\ n_{\mu\nu}\ n_{\alpha\beta}
\ g^{\mu\alpha}\ g^{\nu\beta}}\nonumber \\
&=&\int_S du dv \ \sqrt{g\ (n^T_{\mu}n^\Sigma_{\nu}n^T_{\alpha}
n^\Sigma_{\beta}- \ 
n^T_{\mu}n^\Sigma_{\nu}n^\Sigma_{\alpha}n^T_{\beta})
\ g^{\mu\alpha} g^{\nu\beta}}\nonumber \\
&\equiv&\int_S du dv \ \sqrt{g\ |n^T|^2\
|n^\Sigma|^2-(n^T\cdot n^\Sigma)^2}
\label{anorm}
\label{area}
\end{eqnarray}
This expression gives us the area directly as a function of surface
$S$, the metric and the normals to the table worldsheet and the
observer's simultaneity surface.

\section{Dynamics}\label{geometry}

\subsection{Coordinate choice}\label{coordinates}

Without important loss of generality, we chose coordinates $x^\mu =
(t, x^1, x^2, x^3)$ which are particularly convenient for the above
setting.  In these preferred coordinates: (i) $P$ is the origin.  (ii)
The 3-surface $T$ is defined by $-1<x^1<+1$, $-1<x^2<+1$ and $x^3=0$.
(iii) The world line $X$ is defined by $x^1=x^2=x^3=0$.  (iv) The
world line $X'$ is defined by $x^2=x^3=0$ and $x^1=\beta x^0$.
Furthermore we choose the parameters parametrizing the world lines and
the world sheet as $\tau=\tau'=\tau^3=t$, $u=\tau^1=x^1$ and
$v=\tau^2=x^2$.

We also further fix the coordinates by choosing the gauge in which at
$t=0$ we have $g_{00}=-1$ and $g_{a0}=0$ for $a=1,2,3$.  This
simplifies the canonical analysis.

A very important observation follows.  With these coordinates, namely
in this gauge, the only remaining degrees of freedom are the ones in
the metric tensor $g_{\mu\nu}(x)$.  However, this does not mean that
the physical degrees of freedom of the matter (the two particles and
the table) are being killed or frozen.  Indeed, it is well known that
the coordinate position of matter and the value of the metric tensor
are both gauge dependent quantities, due to diffeomorphism invariance.
The physical, measurable, position of an object (relative to a
reference object) is determined by a combinations of the two.

Let us illustrate this key point in a simple one-dimensional universe
with a metric field $g(x)$, an object $A$ in the coordinate positions
$x=X$ and a reference object in the coordinate position $x=0$.  The
position of $A$ is determined by the distance from the reference
\begin{equation}
d[g,X]=\int_{0}^X dx\ \sqrt{g(x)}.
\end{equation}
$d$ is a diffeomorphism invariant quantity.  It corresponds to what we
actually measure: pick a meter a get the position of $A$ in meters
from the reference -- this measures $d$.  Now, we can gauge fix the
coordinate $x$ so that $g=1$.  In this gauge, the observable quantity
$d$ is given by the coordinate position of the object:
\begin{equation}
    d = X.
\end{equation}
This is what we generally do in flat space: coordinates give
observable positions.  Alternatively, we can choose coordinates in
which the position of the object has a fixed predetermined coordinate
value $X=1$.  In this gauge
\begin{equation}
d =\int_{0}^1 dx\ \sqrt{g(x)}.
\label{distance}
\end{equation}
With this gauge choice, the physical location of the object, namely
its distance $d$ from the reference is determined by the sole metric
field.

These two possibilities are familiar, for instance, in the context of
gravitational wave detectors: We can equivalently say that ``the two
mass probes do not move and the gravitational field varies in the in
between region"; or that ``the two mass probes oscillate in space"
(where ``move" and ``oscillate" refer to the coordinates.)  These are
two equivalent descriptions of the same physics.

In the present context, we have chosen to attach coordinates to the
matter (particles and table).  Therefore the dynamics is entirely
captured by the value of the gravitational field.

To further illustrate how the physical degrees of freedom of the table
and the particle are still present, as well as for later purposes,
consider for instance the following value of the metric, in the given
coordinates:
\begin{eqnarray}
ds^2&=&g_{\mu\nu}(x)\,dx^\mu dx^\nu
 \nonumber \\
 &=& -dt^2 + (1+4\alpha x^1t)
(dx^1)^2+(dx^2)^2 +(dx^3)^2.
\end{eqnarray}
Let $\vec x(t)$ be the trajectory of the central point of this table,
that is, the point at the same distance from its boundaries.
Easily, to first order in $t$,
\begin{equation}
    \vec x(t)= \left(\alpha t,\ 0,\ 0\right).
\end{equation}
Therefore the relative velocity of the particle $X$ with respect to
the center of the table is $\alpha$.  In other words, the particle $X$
and the table $T$ are moving with respect to each other even is their
coordinate positions have been fixed.  The relative velocity of table
and particle is given by the time derivative of the metric field.
($g_{\mu\nu}(x)$ does not depend on $\alpha$ at $t=0$, while its time
derivative does.)  The example illustrate how the physical motion of
the particles and table is described by components of the metric
tensor in this gauge.

Now, since we have partially fixed the gauge by fixing the coordinate
position of the matter, it follows that the invariance under general
coordinate transformations is reduced to the invariance under the
change of coordinates that preserve the coordinate condition chosen.
Equivalently, the diffeomorphisms group {\em Diff}, which is the gauge
group, is reduced in this gauge to the subgroup {\em Diff}${}_{0}$
formed by the diffeomorphisms that send the table and particles'
worldlines into themselves.

As a consequence, certain components of the gravitational field that
are gauge dependent quantities in pure general relativity, become
gauge invariant physical quantities, precisely as the r.h.s.\ of
\Ref{distance}.  In particular, in this gauge the areas $A$ and $A'$,
which are gauge invariant observables, are expressed solely in terms
of $g$, but they still remain, of course, gauge invariant.

We can clarify this point with an analogy from Maxwell theory: in the
gauge in which scalar potential is set to zero, $A_{0}=0$, the
electric field (a gauge invariant quantity) is given by the sole time
derivative of the Maxwell vector potential: $\vec E=d\vec A/dt$.  In
this gauge $d\vec A/dt$ represents a gauge invariant quantity, because
the gauge transformation are reduced to the ones that preserve
$A_{0}=0$.  Similarly, in the coordinates we have chosen the area is
given by a function of $g$ alone, and is gauge invariant because it is
invariant under coordinate transformations that preserve the
coordinate choice made.

We have discussed these issues in great detail in this Section,
because they are sources of frequent confusion.  Let us now write $A$
explicitly as a function of the metric field in the coordinates we
have chosen.

\subsection{Area as function of canonical coordinates}

In the coordinates we have chosen, the table worldsheet $T$ is given 
by
\begin{eqnarray} \label{T}
    T: \left\{
    \begin{array}{c}
    x^3=0,\\
-1 < x^1<+1, \\ -1<x^2<+1
\end{array}
\right.
\end{eqnarray}
and the simultaneity surface $\Sigma$ of the first observer by
\begin{eqnarray}
    \label{Sig}
    \Sigma: & & t=0.
\end{eqnarray}
Therefore
\begin{eqnarray}
n^{\Sigma}_{\mu}=(-1,0,0,0), \nonumber \\
n^T_{\mu}=(0,0,0,1).
\end{eqnarray}
Also, the proper time of the observer coincides with $t$.
Equation \Ref{area} becomes the well known formula
\begin{eqnarray}
\label{areapropria}
A=A(S)=\int_S dudv\ \sqrt{\ \tilde{\tilde g}{}^{33}\ }.
\end{eqnarray}
where we have defined $\tilde{\tilde g}{}^{33}=(-\det g)\ g^{33}$.
Explicitly, since $S$ is the intersection of the surface $\Sigma$
and the table worldsheet $T$  we have from \Ref{T} and \Ref{Sig}
\begin{eqnarray}
\label{areapropria2}
A=\int_{-1}^1 du  \int_{-1}^1 dv \
\sqrt{\ \tilde{\tilde g}{}^{33}(0,u,v,0)}.
\end{eqnarray}

Consider now the observer ${\cal O}'$.  His
simultaneity surface $\Sigma'$ is determined by the worldline $X'$.
The 4-velocity of this world line $X'$ at $P$ is
\begin{eqnarray}
\label{quadrivelocitˆ} v'{}^{\mu}= (1,\ \beta,\ 0,\ 0).
\end{eqnarray}
If $g_{\mu\nu}$ is constant, $\Sigma'$ is just normal to the
4-velocity \Ref{quadrivelocitˆ}: in the parametrization
$\vec\rho=(\rho^1,\rho^2,\rho^3)$ that we have chosen, it is given by
\begin{eqnarray}
    \label{sigmabeta} x^\mu(\vec\rho)=
    (\beta\ g_{1a} \rho^a,\vec\rho)
\end{eqnarray}
where $a=1,2,3$.
Using \Ref{normale}, we have, in the coordinates and parametrization 
chosen
\begin{eqnarray}
    n_{\mu}^{\Sigma'}=(-1,\ \beta g_{1a}),
\label{nmu}
\end{eqnarray}
Since $g_{\mu\nu}(x)$ is in general not constant, the detailed
calculation of the position of $\Sigma'$ is more cumbersome.  We can
shortcut it, to linear order around $P$, by simply taking the value of
$g_{\mu\nu}(x)$ in \Ref{sigmabeta} at a point $\hat x^\mu=\frac{1}{2}
x^\mu$ half way between $P$ and the point of the surface.  (This is in
``the middle" of path of the light that defines $\Sigma'$.  A more
detailed calculation --which we do not report here-- obtained
integrating explicitly the light paths in a metric that grows linearly
in time, confirms the result.)  That is, to next order we replace
\Ref{sigmabeta} by
\begin{eqnarray}
x^\mu(\vec\rho)=
\left(\beta g_{1a}(\hat x(\vec\rho))\,
\rho^a,\vec\rho\right).
\label{x2}
\end{eqnarray}
This equation defines $x(\vec\rho)$ intrinsically, since
$x(\vec\rho)$ appears in the r.h.s. as well. Explicitly, to second
order in $\beta$ we have
\begin{eqnarray}
x^0(\vec\rho)&=&\beta g_{1a}(0,\vec\rho/2)\rho^a
+ \frac{1}{2}\beta^2 \dot g_{1b}(0,\vec\rho/2)\rho^b\ 
g_{1a}(0,\vec\rho/2)\rho^a\
  \nonumber \\
x^a(\vec\rho)&=&\rho^a .
\label{x2exp}
\end{eqnarray}
${\cal O}'$'s simultaneity hypersurface defines the surface
$S^\beta=\Sigma'\cap T$, as the table seen by ${\cal O}'$ at his fixed
time.  (See Figure \ref{3}.) Combining \Ref{x2exp} and
\Ref{T} we have, in the parametrization chosen, again to order 
$\beta^2$
\begin{eqnarray}
x^0(u,v)&=&\beta g_{1i}(u/2,v/2)u^i
 \nonumber \\ &&
+ \frac{\beta^2}{2} \dot g_{1i}(u/2,v/2)u^i  g_{1j}(u/2,v/2)u^j, 
\nonumber \\
x^1(u,v)&=& u^1 = u,  \nonumber \\
x^2(u,v)&=& u^2 = v,  \nonumber \\
x^3(u,v)&=& 0 .
\label{xu}
\end{eqnarray}
Here and from now on, $g_{\mu\nu}(u ,v )=g_{\mu\nu}(0,u ,v ,0)$ and
$g_{\mu\nu}=g_{\mu\nu}(0,0 ,0 ,0)$

From \Ref{area} we have for the second
observer
\begin{eqnarray}
    \label{areaboosted}
A' = A(S')=\int_{S'} dudv\ \sqrt{\ \tilde{\tilde g}{}^{33}\
[1-\beta^2\, g_{11}]}.
\end{eqnarray}
Explicitly,
\begin{eqnarray}
    \label{areaboosted2}
A'=\int_{-1}^1\!\!\!\! du \!  \int_{-1}^1\!\!\!\!  dv \
\sqrt{\ \tilde{\tilde g}{}^{33}(x(u,v))\ [1-\beta^2\, g_{11}(x(u,v)]}.
\end{eqnarray}
Using \Ref{xu}, we have to order $\beta$
\begin{eqnarray}
    \label{areaboostedbeta}
A'&=&A +\frac{\beta}{2} \int_{-1}^1\!\!\!\! du   \int_{-1}^1\!\!\!\!  
dv\
 g_{1i}(u,v)\ u^i\  \nonumber \\ && \ \ \ \ \ \ \ \ \ \times\ \
(\tilde{\tilde g}{}^{33}(u ,v ))^{-1/2}\
\dot{\tilde{\tilde g}}{}^{33}(u ,v ).
\end{eqnarray}
To order $\beta^2$ we have
\begin{eqnarray}
    \label{areaboostedbeta2}
&&\hspace{-.2mm}A'= \int_{-1}^1\!\!\!\! du   \int_{-1}^1\!\!\!\!  dv \
\sqrt{\ \tilde{\tilde g}{}^{33}(x(u,v))} \nonumber \\
&&+\frac{\beta}{2}\int_{-1}^1\!\!\!\! du  \int_{-1}^1\!\!\!\! dv\
\ g_{1i}(x(u,v))\ u^i\
(\tilde{\tilde g}{}^{33}(u ,v ))^{-\frac{1}{2}}\
\dot{\tilde{\tilde g}}{}^{33}(u ,v )\nonumber \\
&&+\frac{\beta^2}{4}
\int_{-1}^1\!\!\!\! du   \int_{-1}^1\!\!\!\!  dv  \ \
 g_{1i}(u,v) u^i\
g_{1j}(u,v) u^j \  \nonumber \\ && \ \ \ \ \ \ \ \ \ \times\ \
(\tilde{\tilde g}{}^{33}(u ,v ))^{-\frac{1}{2}}\
\ddot{\tilde{\tilde g}}{}^{33}(u ,v )
\nonumber \\
&&-\frac{\beta^2}{8} \int_{-1}^1\!\!\!\! du  \int_{-1}^1\!\!\!\!  dv  
\
(\tilde{\tilde g}{}^{33}(u ,v ))^{-\frac{3}{2}}\
(g_{1i}(u,v)u^i)^2
(\dot{\tilde{\tilde g}}{}^{33})^2(u ,v )
\nonumber \\
&&-\frac{\beta^2}{2} \int_{-1}^1\!\!\!\! du  \int_{-1}^1\!\!\!\!  dv \
g_{11}(u,v)\ (\tilde{\tilde g}{}^{33}(u ,v ))^{\frac{1}{2}},
\end{eqnarray}
and so on. The second, third and fourth line of this equation come
from the time derivatives of $\tilde{\tilde g}{}^{33}$, which depend
on $\beta$. The last line comes from the $\beta^2$ in
\Ref{areaboosted2}. Notice that there is still a $\beta$ dependences
in the first two lines, because $x(u,v)$, given in \Ref{xu}, contains
$\beta$.

Let us now make the additional assumption that the metric is spatially
constant at $t=0$.  This simplifies the expressions above and allows
us to perform the integrals explicitly, but it is not essential: it is
easy to generalize our result to a non spatially constant metric.
Under this condition we can write
\begin{eqnarray}
g_{\mu\nu}(x(u ,v ))&=&
g_{\mu\nu}+\beta\ g_{1i} u^i \ \dot g_{\mu\nu}+
\frac{\beta^2}{2} g_{1j}u^j \ \dot g_{1i} u^i \  \dot g_{\mu\nu}
\nonumber \\
&&+\frac{\beta^2}{2} g_{1j}u^j \  g_{1i} u^i \  \ddot g_{\mu\nu}.
\end{eqnarray}
Inserting this in the first two lines of \Ref{areaboostedbeta2}, we
can do all the integrals explicitly.  The ones linear in $u$ vanish,
leaving, to second order in $\beta$, with a straightforward
calculation
\begin{eqnarray}
    \label{aree}
    A &=& \sqrt{\tilde{\tilde g}{}^{33}} \\
    \label{aree2}
    A'&=&A-2
    \beta^2\ g_{11} A
\nonumber \\
&&  + \beta^2\  (g_{11}^2 + g_{22}^2 )  \left(
\frac{\partial^2_{t}\, \tilde{\tilde g}{}^{33}}{3A}
- \frac{(\partial_{t}\, \tilde{\tilde g}{}^{33})^2 }{6A^3}\right)
 \nonumber \\
&& + \ \beta^2\
    (g_{11}\dot g_{11}+g_{12}\dot g_{12})\
\frac{\partial_{t}\, \tilde{\tilde g}{}^{33}}{3A}.
\end{eqnarray}

\section{Non commutativity}\label{incomp}

So far, we have simply studied the form of the areas $A$ and $A'$ seen
by two accelerated observers, in a given metric.  Let us now recall
that the metric is the gravitational field, namely a dynamical
physical field.  We want to write $A$ and $A'$ as functions on the
phase space of our dynamical theory, and compute the Poisson bracket
$\{ A,A' \}$.  To this aim, we take the simultaneity surface $\Sigma$,
that is $t=0$ in the coordinates chosen, as our $ADM$ surface, on
which we base the canonical formalism.  As usual in quantum gravity,
we chose as canonical variable the Ashtekar's field $E^a_{i}(x)$,
namely the densitized tetrad field, which satisfies
\begin{eqnarray}
E^a_{i}(x)E^b_{i}(x)= \tilde{\tilde g}{}^{ab}(x).
\label{ee}
\end{eqnarray}
We consider the metric field (which we leave indicated when
convenient) as a function of the tetrad field.
The explicit form of the brackets in the r.h.s.\ depends on the
dynamics of the matter field, which in the coordinates we have chosen
affects the dynamics of the gravitational field by partially
constraining the evolution of Lapse and Shift.  However, one can
easily see that even if we assume that in these coordinates the
dynamics of the gravitational field is unaffected by the matter, the
r.h.s.\ does not vanish.  In this case, indeed, we can take the
evolution in the coordinate $t$ to be generated simply by the pure
gravity hamiltonian constraint (we are in Lapse=1, Shift = 0 gauge),
namely
\begin{eqnarray}
\dot E_{i}^a &=&  \{E^a_{i},H \},\nonumber \\
\dot A_{a}^i &=&  \{A^i_{a},H \},
\end{eqnarray}
where
\begin{eqnarray}
H&=&\int d^3x \ E_{i}^a E_{j}^b F_{ab}^k \epsilon_{k}{}^{ij}\nonumber 
\\
&=&\int d^3x \  E_{i}^a E_{j}^b\ (\partial_{a}A^k_{b} 
\epsilon_{k}{}^{ij}
+A^i_{[a}A^j_{b]})).
\end{eqnarray}
Recalling that the non vanishing Poisson brackets are given by
\begin{eqnarray}
\{E^a_{j}(x),A^k_{b}(y)\}=\delta^a_{b}\delta^k_{j}\ \delta^3(x,y),
\end{eqnarray}
we can compute the Poisson brackets explicitly.

Surprisingly,
\begin{eqnarray}
\{\tilde{\tilde g}{}^{33},\dot{\tilde{\tilde g}}{}^{33}\}=0
\label{undot}
\end{eqnarray}
and
\begin{eqnarray}
\{\tilde{\tilde g}{}^{33},\ddot{\tilde{\tilde g}}{}^{33}\}=0.
\label{duedot}
\end{eqnarray}
The first equality follows from
\begin{eqnarray}
\{\tilde{\tilde g}{}^{33},\dot{\tilde{\tilde g}}{}^{33}\}&=& 2
E^3_{i}E^3_{j}\{E^3_{i},\dot E^3_{j}\} \nonumber \\ &=& 2
E^3_{i}E^3_{j}\{E^3_{i},E^3_{[j} E^a_{k]} A_{a}^k \} \nonumber \\
&=& 2 E^3_{i}E^3_{j}\ E^3_{[j} E^3_{i]}\nonumber \\
&=& 0.
\end{eqnarray}
The second equality can be derived from
\begin{eqnarray}
\{\tilde{\tilde g}{}^{33},\ddot{\tilde{\tilde g}}{}^{33}\}=
E^3_{i}\{E^3_{i},(2\dot E^3_{j}\dot E^3_{j}+E^3_{j}\ddot E^3_{j})\}
\end{eqnarray}
Of the two terms on the r.h.s., the first is proportional to
$\{E^3_{i},\dot E^3_{j}\}$, which, as he have just seen, vanishes.
The second can be written as
\begin{eqnarray}
E^3_{i}E^3_{j}\{E^3_{i},\ddot E^3_{j}\} = E^3_{i}E^3_{j}
(\p_{t}\{E^3_{i},\dot E^3_{j}\}-\{\dot E^3_{i},\dot E^3_{j}\});
\end{eqnarray}
again, the first term in the parenthesis vanishes as we have seen,
while second term is antisymmetric in $i$ and $j$.  Thus, only the
last term of \Ref{aree2} does not commute with \Ref{aree}. A
long but straightforward calculation gives indeed
\begin{eqnarray}
\{A,A'\} = \frac{8}{3} \beta^2\ (g_{11}^2+g_{12}^2) \ A\dot A.
\end{eqnarray}
Since the Poisson brackets between $A$ and $A'$ do not vanish, the
commutator of the corresponding quantum operator cannot vanish either.
Otherwise in the $\hbar\to 0$ limit the commutator could not reproduce
the classical Poisson brackets.  This confirms our main claim,
equation \Ref{eq:nc}.

For later purposes we write also the expression for $A'$ to
first order in $\beta$ without the assumption of spatially constant
metric: inserting \Ref{ee} in \Ref{areaboostedbeta} we get
\begin{eqnarray}
    \label{areaboostedbetatetr}
A'=A&+&\beta \int_{-1}^1\! \! du  \int_{-1}^1\! \!  dv\  g_{1i}(u,v) 
u^i\
\ \ \times \nonumber \\ && \sqrt{E^3_{i}(u ,v )E^3_{i}(u ,v )} \
E^3_{i}(u ,v )\dot E^3_{i}(u ,v ).
\end{eqnarray}

\subsection{The velocity of the surface}\label{ac}

We close the chapter with an observation, that we learned in
discussions with Amelino-Camelia.  In flat space, the area $A'$
observed by ${\cal O}'$ is related to the area $A$ seen by an observer
$\cal O$ at rest with respect to the surface by
\begin{eqnarray}
A' \ = \ \sqrt{1-{V^2}}\ \ A \label{ApgA}
\end{eqnarray}
where $V$ is the relative velocity of the two observers.  If the
surface is sufficiently small, the same should be true in general
relativity.  But this seems in contradiction with (\ref{eq:nc}), which
we claim to be the key to understand the problem at hand.  Indeed, if
$A$ is an operator, (\ref{ApgA}) seems to express $A'$ as a simple
function of $A$: but a function of an operator commutes with the
operator itself, therefore $A'$ should commute with $A$, against
(\ref{eq:nc}).  The answer to this objection is illuminating.  In
general relativity, $A\,$ becomes a quantum operator because it is a
function of the metric, namely a function of the quantum field.  But
the velocity $V$ that appears in (\ref{ApgA}) depends on the metric as
well.  Indeed the $V$ that appears in (\ref{ApgA}) is not a coordinate
velocity, it is a physical velocity, and it depends on $g_{\mu\nu}$ as
well.  Thus $V$ as well is an operator in quantum gravity.  Therefore
the operator $A'$ is not a simple function of the operator $A$.  The
non commutativity (\ref{eq:nc}) of $A$ and $A'$ can thus be
equivalently viewed as a consequence of the non commutativity of $A$
and $V$.  Therefore one can also say that the apparent incompatibility
between discreteness and Lorentz contraction is resolved by observing
that the measurements of area and velocity of a surface are
incompatible.

To be more precise, since the relative velocity between observer and
surface does not commute with the area, it does not make sense to
start by assuming that the first observer $\cal O$ is at rest with
respect to the surface. Dropping this, we must replace (\ref{ApgA}) by
\begin{eqnarray}
A' \ = \ \frac{\sqrt{1-{v'{}^2}}}{\sqrt{1-{v^2}}}\ \ A
\label{ApgA2}
\end{eqnarray}
where $v$ and $v'$ are the relative velocities of the two observers
$\cal O$ and ${\cal O}'$ with respect to the surface.  In general
relativity this becomes complicated because the notion of rest frame
of a non-local object --such as the surface-- is far more complicated
than in special relativity.  Since the location of the surface we
consider is only defined by its boundary, its rest frame depends only
on its boundary as well.  The distance of the boundary from the
observer is determined by the value of the gravitational field on the
surface itself; the velocity of the boundary with respect to the
observer (that is, the rate of change of this distance in the
observer's proper time) depends, therefore, on the \emph{time
derivative} of the gravitational field.  This is shown above in a
concrete example in Section \ref{coordinates}.  Since the
gravitational field operator does not commute with its own time
derivative, this velocity does not commute with the area.  As $v$ and
$v'$ do not commute with $A$, $A'$ does not commute with $A$ either.
This point is discussed in detail by Amelino-Camelia in
\cite{giovanni}.

Physically, all this means that by measuring the area, an observer
destroys information on the velocity of the surface ---as measuring
the position of a quantum particle destroys information on its
momentum.

In facts, one might have considered another possible solution for the
apparent conflict between Lorentz contraction and discreteness.
Recall that one can say that the rest energy $E_{0}$ of a massive
particle is a non-Lorentz-invariant quantity (it is the fourth
component of a four-vector), but it is also a fixed fundamental
observable quantity in a Lorentz invariant theory.  There is no
contradiction, because $E_{0}$ is measured in a special frame
determined by the state itself.  Similarly, we might imagine that
$A_{0}$ always appears as the minimal area of a material object in its
own rest frame.  The explicit computation of this paper shows that
this is not the case.  But the observation above clarifies why: a
measurement of the area erases information on the velocity of the
surface.  Presumably, a quantum measurement of the area $A$ projects
the system into a state in which $v$ is maximally spread: then the
mean value of this velocity is in any case zero \emph{after} the
measurement.

\section{Boosts generators}\label{quantum}

We now want to study the transformation that maps the operators $A$
and $A'$, corresponding to the classical quantities $A$ and $A'$, into
each other.  In particular, we are interested in understanding if this
transformation can be seen, in an appropriate sense, as a Lorentz
transformation.  The subtlety is the interplay between the assumption
of approximate local flatness of the mean values of the quantum fields
and the full dynamical structure of the theory.  We place ourselves in
the frame of the full theory, but studied in the vicinity of the
states which are macroscopically flat around $P$.  We suggest here
that in this context one can define a unitary transformation in the
Hilbert space of the theory, which sends $A$ into $A'$.  If this is
correct, the spectrum of the two operators is the same, a result which
is to be expected on physical grounds.

To this aim, we explicitly consider quantities $M^\mu{}_\nu$ that
behave as generators of Lorentz transformations.  For a field theory
on flat space, the construction of these quantities is well known (see
for instance \cite{230}).  We briefly recall it here.  Define
\begin{equation}
\label{generatori} M^\mu{}_\nu=\int{d^3x}\left[
x^{[\mu}T^{0|\rho]}\ \eta_{\rho\nu}-i\frac{\p{\cal{L}}}
{\p\dot{\phi}^n}(L^\mu{}_\nu)^n_m \phi^m \right],
\end{equation}
where we have indicated by $\phi^m(x)$ the fields; $m$ is a
generic Lorentz index; $T^{\mu\nu}$ is the energy momentum
tensor, ${\cal{L}}$ is the Lagrangian density (and therefore
$\frac{\p{\cal{L}}}{\p\dot{\phi}^n}$ are the momenta conjugate to
the fields), $(L^\mu{}_\nu)^n_m$ are the generators of the
Lorentz representation to which the fields belong, and
$\eta_{\mu\nu}$ is the Minkowski metric.  In a Lorentz invariant
theory, \Ref{generatori} are constant.  Let us indicate by
$q^n(x)$ the canonical fields, by $p_n(x)$ their conjugate
momenta.  Other fields will be the auxiliary ones --the ones with
vanishing conjugate momentum.  We can write
\begin{equation}
    M^a{}_{b}=\int{d^3x}\ p_n(x)\left[
x^{[c}\p^{a]}q^n(x)\ \eta_{cb}-i(L^a{}_{b})^n_m q^m(x) \right].
\end{equation}
and it is easy to verify that these are indeed generators of spatial
rotations.  More care is required for the boosts, because in general
they mix canonical and auxiliary fields:
\begin{eqnarray}
    \label{boost}
M^0{}_{a}&=&\int{d^3x}\left[-x^0 p_n(x) \p_a q^n(x)-\eta_{ac} x^c
T^{00}(x)\right. \nonumber \\ &&
\left. -ip_n(x)(L^0{}_{a})^n_m \phi^m(x) \right]
\end{eqnarray}
[Notice that these quantities are constants in time, but they do not
commute with the Hamiltonian --in fact, they Lorentz transform the
Hamiltonian $H$ into the total momentum $P_{a}$, as is to be expected
geometrically.  This is because of they are explicitly time
dependent:
\begin{equation}
0=\dot{M}^0{}_{a}=\left\{H,M^0{}_{a}\right\}+\frac{\p M^0{}_{a}}{\p t}
\end{equation}
from which
\begin{equation}
    \left\{H,M^0{}_{a}\right\}=-\int{d^3x} \ p_n(x)
\p_a\ q^n(x)=P_a.
\end{equation}
This is why they do not give good quantum numbers in spite of being
constant.]

Let us now come to gravity.  In gravity, we can still write the
quantities \Ref{generatori}.  These are formal objects.  They are not
tensorial, not defined for all values of the fields, not defined on
the entire spacelike surface.  Nevertheless, they can still play a
role.  Indeed, let us consider the transformation they generate over a
function of the fields which has support in a region small with
respect to the local curvature, or in a regime in which spacetime is
close to flatness.  In this regime, we can take the $x^\mu$ as
cartesian coordinates, and we can take these objects as the generators
of Lorentz transformations.

Consider in particular the component $T^{00}(x)$.  This is the
hamiltonian constraint density, since
\begin{equation}
    \label{hamdensity}
T^{00}(x)=H(x)={\sum}_m\frac{\p {\cal{L}}}{\p
\dot{\phi}^m}(x)\dot{\phi}^m(x)-{\cal{L}}(x).
\end{equation}
If we can fix the gauge $N(x)=1$, $N^a(x)=A_0^i(x)=0$,
$\dot{A}=\left\{A,{\cal{H}}\right\}$, the hamiltonian density
\Ref{hamdensity} coincides with the hamiltonian constraint ${\cal
H}(x)$ \cite{232}.  The momentum,
\begin{equation}
    \label{tzeroa} T^0{}_{a}(x)=- \frac{\p
{\cal{L}}}{\p \dot{\phi}^m}(x)\p_a \phi^m(x),
\end{equation}
generates spatial translations. Spatial translations are generated
by the momentum constraint ${\cal{H}}_a(x)$ in general
relativity.  At the light of these considerations, we tentatively
consider the possibility that the boost generator that sends the
area in the boosted area is given by
\begin{eqnarray}
\label{boostLQG} M^0{}_{a}&=&\!\int \! {d^3x}\ [x^0
{\cal{H}}_a(x)+x^0 A_{a}^i(x){\cal G}^i(x)
 \nonumber \\ &&
- g_{a\mu}(x) x^\mu{\cal{H}}(x)-i 
E^b_i(x)(L^0{}_{a})_b^{\mu}A^i_{\mu}(x)]
\end{eqnarray}
Notice the replacement of the Minkowski metric by $g_{\mu\nu}(x)$.
More precisely, we consider the possibility that an infinitesimal
Lorentz boost $\lambda^a{}_0$ acting in the point $x$, is generated by
\begin{eqnarray}
\label{boostinfinitesimo}
M(\lambda)=\lambda^a{}_0 \ M^0{}_a
\end{eqnarray}
In our case, $x^0=0$ and
\begin{eqnarray}
\lambda^\mu{}_\nu=\beta\ \epsilon^\mu{}_{\nu 2 3};
\label{lambda}
\end{eqnarray}
therefore the generator turns out to be
\begin{eqnarray}
\label{boostinfinitesimo2}
M(\beta)=\beta\  M^0{}_{1}
\end{eqnarray}
Taking account that we are at $x^0=t=0$ and in the gauge $A^i_{0}=0$,
we have
\begin{eqnarray}
\label{boostinfinitesimo3} M(\beta)=-\ \beta\ \int d^3x \
g_{1a}(x) \ x^a\ {\cal H}(x)
\end{eqnarray}

In order to check this hypothesis, we compute the infinitesimal
transformation of $A$ generated by this generator. Since
\begin{eqnarray}
&& \left\{M(\beta),\sqrt{E^3_j(x)E^3_j(x)} \right\}=
 \nonumber \\ &&\ \ \ =
\int{d^3z}\ \frac{\delta M(\beta)}{\delta A^i_b(z)}\ \frac{\delta
\sqrt{E^3_j(x)E^3_j(x)}}{\delta E_i^b(z)}\nonumber \\  &&\ \ \ =
\frac{\delta M(\beta)}{\delta A^i_3(x)} \
[E^3_j(x)E^3_j(x)]^{-{\frac{1}{2}}}\ E^{3i}(x)\nonumber \\   &&\
\ \ = -\beta \int{d^3y}\  g_{1a}(y) y^a\ \frac{\delta {\cal
H}(y)}{\delta A^i_3(x)} \ [E^3_j(x)E^3_j(x)]^{-{\frac{1}{2}}}\
E^{3i}(x)\nonumber \\  &&\ \ \ = \ \beta g_{1a}(x) x^a\
\dot{E}_i^3(x)\ [E^3_j(x)E^3_j(x)]^{-{\frac{1}{2}}}\ E^{3i}(x),
\end{eqnarray}
it follows
\begin{eqnarray}
\left\{M(\beta),A \right\}&=&
\beta \int_{-1}^1\!\!du  \int_{-1}^1\!\! dv\  g_{1i}(u,v) u^i\
\ \times  \nonumber \\ &&
\sqrt{E^3_{i}(u ,v )E^3_{i}(u ,v )}\
E^3_{i}(u ,v )\dot E^3_{i}(u ,v ).
\end{eqnarray}
But this is precisely the second term in the r.h.s of Equation
(\ref{areaboostedbetatetr}), which is the infinitesimal
transformation of $A$ we had previously worked out
geometrically.  This result supports the hypothesis that
(\ref{boostinfinitesimo2}) is the correct generator of the
local Lorentz boost.

\subsection{Unitarity}\label{unitarity}

Let us now return to the quantum theory.  Consider the quantum
operator $M(\lambda)$ corresponding to the classical observable
\Ref{boostinfinitesimo}.  We assume that this operator is well defined
in the theory.  The corresponding finite transformation is generated
by
\begin{equation}
\label{boosts}
U(\lambda)=e^{-iM(\lambda)},
\end{equation}
This operator is unitary if $M(\lambda)$ is hermitian.  This is the
condition under which the Lorentz transformation is unitary in the
quantum theory.  Assuming it is satisfied, the spectrum of the areas
$A$ and $A'$ is the same.  Conversely, since on physical grounds
nothing distinguishes $A$ from $A'$, we think it is reasonable to
require that the operator $M(\lambda)$ be hermitian.

Let us study this condition.  Consider the infinitesimal action of the
operator (\ref{boosts}) on the states of the theory.  We take $x^0=0$
and $\lambda$ given by \Ref{lambda}, so that \Ref{boostLQG} reduces to
\Ref{boostinfinitesimo3}
\begin{eqnarray}
|\psi_\beta\rangle = |\psi\rangle +i\beta \int{d^3x}\ g_{1a}(x)\
x^a\ {\cal{H}}(x)\ |\psi\rangle
\end{eqnarray}
and is determined by the hamiltonian operator.  Recall that a basis of
area eigenstates is given by spin network states
\cite{spinnet,spinnet2}.  We denote a spin network state as
$|\Gamma,j\rangle$, where $\Gamma$ is a graph, and $j$ the coloring
associated to the links and nodes $n$ in $\Gamma$.  We can expand
\begin{equation}
    |\psi \rangle = \sum_{\Gamma, j} \psi_{\Gamma,j}\ | \Gamma, 
j\rangle
\end{equation}
We recall that the action of the Hamiltonian constraint smeared with a
Lapse $N$ is a sum of terms acting on the nodes of the form
\cite{cc,cc2}
\begin{equation}
\label{H_on_s}
  \hat{H}[N] \, | \Gamma, j \rangle =
    \sum_{n \in \Gamma}
    A_n \, N(x_n)\, {D}_n \, | \Gamma, j \rangle.
\end{equation}
where $x_n$ is the coordinate location of the $n$-th node, $A_{n}$ are
numerical coefficients and ${D}_n$ is an operator that acts on the
graph changing it around the $n$-th node.  See \cite{loop2} and
especially \cite{tom} and references therein, on the actual
construction of the hamiltonian constraint operator.  Here the Lapse
is one.  Using all this we obtain
    \begin{eqnarray}
    |\psi_\beta\rangle
    \simeq
    | \psi \rangle+i\beta
    \sum_{\Gamma, j} \psi_{\Gamma, j}
    \sum_{n \in \Gamma} A_n \, (g_{1a}x^a_n)\, {D}_n \, | \Gamma, j 
\rangle
\end{eqnarray}
In particular, if consider a spin network $| \Gamma, j \rangle$,
eigenstate of $A$, the probability amplitude that ${\cal O}'$ sees it
in a different spin network eigenstate $| \Gamma', j' \rangle$ is
    \begin{eqnarray}
    P = \beta
    \sum_{n \in \Gamma}
    A_n \ \langle \Gamma', j'
    |g_{1a} x^a_n\, {D}_n \, | \Gamma, j \rangle.
    \end{eqnarray}
We leave the problem of the actual definition of the node operator
$g_{1a} x^a_n\, {D}_n$ in the quantum theory to future
investigations.

\section{Discussion and conclusions}

In loop quantum gravity the metric is an operator.  The area of a
surface is a quantum observable.  At the Planck scale, this area is
quantized and there is a finite nonzero minimal value.  Under a
Lorentz transformation, we expect this minimal value not to change.
That is, we expect that two observers boosted with respect to each
other, see the same spectrum.  We have studied here the transformation
that relates the observables of the two observers.

We have analyzed in detail the situation in classical general
relativity, and written the form of the two observables explicitly.
We have shown that these two observables have nonvanishing Poisson
brackets, which implies that the corresponding quantum operators
cannot commute.  Therefore if the value of the area is sharp for one
observer, it cannot be sharp, in general, for the second observer.
This implies that the minimal area measured by one observer cannot be
just Lorentz contracted for the boosted observer. This is our main
result.

We have also studied the conditions under which the transformation
between the two observables is unitary in quantum theory.  These
conditions can be seen as requirement on the precise definition of
certain operators in the quantum theory.  We have suggested the
explicit form of the generator of local Lorentz transformations in the
theory, in a particular gauge.

We close with a discussion of the relation between diffeomorphism
invariance and Lorentz transformations, in this context.  The theory
is invariant under diffeomorphisms that act simultaneously on the
gravitational field and on the matter.  However, it is not invariant
under a diffeomorphism that acts on the matter leaving the
gravitational field untouched.  Nor under a diffeomorphism that acts
on the gravitational field leaving the matter untouched.  Of course,
diffeomorphism invariance implies that to move the matter with respect
to the gravitational field is equivalent to moving the gravitational
field with respect to the matter.  The Lorentz transformations we have
considered act on the matter at fixed field, or, equivalently, on the
field leaving the matter fixed.  This is why they are not part of the
gauge.  Concretely, we have gauge fixed the coordinate position of the
matter, and considered an active Lorentz transformation rotating (in
spacetime) the gravitational field.  While this would be a gauge
transformation in the absence of matter and in arbitrary coordinates,
it is, instead, a change of physical state in the presence of matter,
or, equivalently, in the gauge fixed coordinates we have chosen.  This
is why, in spite of being a linear function of the hamiltonian
constraint, the generator of Lorentz transformations that we have
introduced defines a genuine transformation in the physical Hilbert of
the theory.  Technically, since we have gauge fixed the coordinates,
the physical states are not defined by the vanishing of the full
constraints, but only by the vanishing of the constraints smeared by
generators of diffeomorphisms that send the matter worldhistories into
themselves.

At the light of these considerations, the reason for the explicit form
of the generator we have considered (see in particular
\Ref{boostinfinitesimo3}) is transparent: it changes the value of the
metric field from the one on the $t=0$ surface to the one of the
surface $t=\beta g_{1a} x^a$, namely to the Lorentz rotated surface. 
Therefore it transforms the gravitational field that determines the
area of the table on the simultaneity surface of the first observer
into the field that determines the area of the table on the
simultaneity surface of the boosted observer.

\end{document}